\documentclass[aps,preprint]{revtex4}
\def\beq{\begin{equation}}
\def\eeq{\end{equation}}
\usepackage{graphics,epsfig,amssymb}

\begin{document}

\title{Isovector Neutron-Proton Pairing with Particle Number Projected BCS}

\author{N. Sandulescu}
\affiliation{
Institute of Physics and Nuclear Engineering, P.O. Box MG-6, 76900 Bucharest-Magurele, Romania}
\author{ B. Errea and J. Dukelsky}
\affiliation{
Instituto de Estructura de la Materia, CSIC, Serrano 123, 28006 Madrid, Spain}
\vskip 3cm
\begin{abstract}
The particle number projected BCS (PBCS) approximation is tested against the exact solution of the SO(5)
Richardson-Gaudin model for isovector pairing in a system of non-degenerate single particle orbits. Two isovector
PBCS wave functions are considered. One is constructed as a single proton-neutron pair condensate, while the other
corresponds to a product of a neutron pair condensate and a proton pair condensate. The PBCS equations are solved
using a recurrence method and the analysis is performed for systems with an equal number of neutrons and protons
distributed in a sequence of equally spaced 4-fold (spin-isospin) degenerate levels. The results show that
although PBCS improves significantly over BCS, the agreement of PBCS with the exact solution is less
satisfactory than in the case of the SU(2) Richardson model for pairing between like particles.

\end{abstract}

\maketitle

\section{introduction}

Neutron-proton ($np$) pairing is a longstanding issue in nuclear structure \cite{lane}. Despite many efforts, the
specific fingerprints of these correlations in existing nuclear data are not yet clear, nor the appropriate
theoretical tools for their correct treatment. For many years the theoretical framework commonly used to describe
the $np$ pairing correlations was the generalized HFB approach \cite{goodman1}. In this approach the $np$ pairing,
both isovector and isoscalar, is treated simultaneously with neutron-neutron ($nn$) and proton-proton ($pp$)
pairing. However, although the generalized BCS approach treats on equal footing all type of pairing correlations,
most of BCS calculations show that they rarely mix \cite{goodman2}. Thus, in general, there are three BCS
solutions which seem to exclude each other: one with $nn$ and $pp$ pairs; the second, degenerate to the first in
even-even $N=Z$ nuclei, with isovector $np$ pairs; and the third with isoscalar $np$ pairs.

Various studies have shown that the restoration of particle and isospin symmetries and the inclusion of higher
order correlations improve significantly the predictions of BCS approach for systems with $np$ pairing
\cite{chen,engel,satula,dobes,delion}. To restore exactly these symmetries, projection operators or projected
generator coordinate methods are commonly employed \cite{ring_schuck}. Less discussed in the literature is an
alternative method based on the recurrence relations satisfied by the isovector pairing Hamiltonian averaged on
projected BCS (PBCS) wave functions. In this paper we will implement this method to analyze the dependence of
isovector pairing correlations on particle number conservation. As trial wave functions we will use two PBCS
condensates, one formed by isovector $np$ pairs and another by $nn$ and $pp$ pairs. Contrary to the BCS
approximation for a system of an even number of pairs, the PBCS solutions corresponding to these two pair
condensates are not degenerate. To analyze how much these PBCS solutions could improve over the generalized BCS
approach will shall use the exactly solvable SO(5) Richardson-Gaudin pairing model \cite{so5}. Several previous
studies have been carried out in the one-level degenerate SO(5) model \cite{Hecht}. These studies clarified the
limitations of the BCS approximation, and the corresponding extensions taking into account pair fluctuations in
the RPA formalism or using boson expansion theories \cite{engel,delion2,dobes}.  Studies on number and isospin
projection on the isovector pairing Hamiltonian with non-degenerate single-particle levels have been reported in
\cite{chen}. However, these studies were tested against a solution proposed by Richardson \cite{richardson_so5}
and later on shown to be incorrect for systems with more than two pairs \cite{pan}. The exact solution of the
non-degenerate isovector pairing Hamiltonian has been given by Links {\it et al.} \cite{links} and afterwards
generalized to seniority non-zero states, arbitrary degeneracies, and symmetry breaking Hamiltonians in
\cite{so5}. This solution will be used here as a benchmark to test the accuracy of PBCS approximations for
describing the isovector pairing correlations.

\section{Formalism}

We will consider an isovector ($T=1$) pairing Hamiltonian with a constant pairing strength
\begin{equation}
\hat{H}= \sum_{im\tau} \varepsilon_{j_i} a^{\dagger}_{j_im\tau} a_{j_im\tau}- g \sum_{i,i',\tau}
\sqrt{(j_{i}+1/2)(j_{i'}+1/2)} P^+_{j_{i}\tau} P_{j_{i'}\tau},
\end{equation}
where  $P^+_{j_i} =\frac{1}{\sqrt{2}}[a^+_{j_i} a^+_{j_i}]^{01}_{0\tau}$ is the isovector pair creation operator. The first
column in the couplings refers to total angular momentum and the second column to total isospin.

The Hamiltonian (1) is a particular example of the exactly solvable SO(5) Richardson-Gaudin integrable models. It is
is exactly solvable for arbitrary single particle energies $\varepsilon_{j_i}$ and pair degeneracies $j_i+1/2$.
The exact solution of these class of Hamiltonians has been given in Ref. \cite{so5}. Here we will treat
a simplified version for a system of $L$ equidistant single- particle levels of pair degeneracy 1,
that is $j_i=1/2$. The exact solution for this system  will be used as a  test for the PBCS approximation with 
isovector pairing.
For comparison we shall also show the results of the proton-neutron BCS approximation. The generalized BCS model
used in this paper is described in Ref. \cite{bes}. As in the case of a single degenerate level \cite{dobes},
within the BCS approximation the Hamiltonian (1) has two solutions: ({\it A}) a BCS solution with a non-zero
proton-neutron gap, $\Delta_{np} \neq 0$, and zero gaps for neutron-neutron and proton-proton pairs, i.e.,
$\Delta_n=\Delta_p=0$; ({\it B} ) a BCS solution with $\Delta_n=\Delta_p \neq 0$ and $\Delta_{pn}=0$. The two
solutions ({\it A} ) and ({\it B} ) exclude each other and are degenerate in energy for a system with an even
number of pairs. In the next section we will present the PBCS equations corresponding to these two BCS solutions.
The PBCS formalism will be given in the form of recurrence relations, and it can be applied to general
(density-independent) isovector pairing interactions, irrespectively of whether they are integrable or not.

\subsection{ PBCS approximation with isovector proton-neutron pairs}
\vskip 0.3cm

We shall first consider a PBCS wave function corresponding to the solution (A), i.e., formed by $N$ isovector
neutron-proton pairs. It has the following form \beq | N > = \frac{1}{N!} (\Gamma^+_0)^N | 0 >, \eeq where
$\Gamma^+_0$ is the collective neutron-proton pair operator \beq \Gamma^+_0 = \sum_{i=1}^{L} x_i P^+_{i0}. \eeq
This wave function is not normalized and the factor in front is chosen to simplify the form of PBCS equations. The
mixing amplitudes $x_i$ are determined
 by minimizing the energy functional
\beq
E(x) = \frac{<N| H | N>}{<N|N>}.
\eeq
The norm and the expectation value of the Hamiltonian are calculated by using
recurrence relations.  Thus, it can be shown that the norm of the wave function (2)
satisfies the equation
\beq
~~<N|N> =  \frac{1}{N} \sum_i x_i^2 <N-1|N-1> - \frac{1}{2N} \sum_i x^3_i <N-1| P^+_{i0}|N-2>
\eeq
where
\beq
<N|P_{i0}^+|N-1> = x_i <N-1|N-1> - \frac{1}{2} x^2_i <N-1| P^+_{i0}|N-2>.
\eeq
To get the norm corresponding to the system with $N$ proton-neutron pairs the equations
above should be iterated starting with $<1|1> = \sum_i x_i^2$ and $<1|P_{i0}^+|0>=x_i$.

The expectation values of the particle number operators $N_i$, which give the occupation probabilities of the
single-particle levels, can be calculated from the equation \beq <N|N_i|N>= 2 x_i <N| P^+_{i0}|N-1> \eeq where the
matrix elements in the r.h.s. are given by Eq.6.

Finally, the matrix elements of the pairing force are given by the equations
\begin{eqnarray*}
<N|P^+_{i0}P_{j0}|N> &= & \frac{1}{4} x_i^2 x_j^2 <N-2|P^+_{j0}P_{i0}|N-2> \\
 &  & + x_j <N|P^+_{i0}|N-1> - \frac{1}{2} x_j^2 x_i <N-1|P_{j0}^+|N-2> \\
 &  & + \delta_{ij} \frac{x^4_i}{4} [<N-2|N-2> - \frac{1}{2} <N-2|N_i|N-2>]
\end{eqnarray*}
\begin{eqnarray*}
<N|P^+_{i1}P_{j1}+P^+_{i-1}P_{j-1}|N> & = &
\frac{x_i^2 x_j^2}{4} <N-2|P^+_{i1}P_{j1}+P^+_{j-1}P_{i-1}|N-2> \\
 &  & + \delta_{ij} \frac{x_i^4}{2} [<N-2||N-2> - \frac{1}{2} <N-2|N_i|N-2>]
\end{eqnarray*}
These equations above are iterated  starting from $<1|P^+_{i0}P_{j0}|1> = x_i x_j$ and
$<1|P^+_{i1}P_{j1}+P^+_{i-1}P_{j-1}|1> = 0$.

\subsection{ PBCS approximation with proton-proton and neutron-neutron pairs}

We will now consider a PBCS wave function corresponding to the BCS solution (B), i.e., given by a product of two
condensates formed by $nn$ and $pp$ pairs. This trial wave function has the form \beq |M M> \equiv |M>\otimes|M> =
\frac{1}{(M!)^2} (\Gamma_n^+ \Gamma_p^+)^{M} |0> \eeq where $M$ denotes the number of $nn$ and $pp$ pairs,
$M=N/2$, while $\Gamma_n^+$ and $\Gamma_p^+$ are the collective pair operators for neutrons and protons (see
Eq.(9) below). As defined here, the wave function (8) is well suited for even-even nuclei. For odd-odd nuclei the
corresponding  wave function is formed by $M=(N-1)/2$ neutron-neutron and proton-proton pairs plus two
unpaired nucleons that block the corresponding levels affecting the pairing correlations.

Since the Hamiltonian (1) is symmetric in isospin, for $N=Z$ systems the collective proton and neutron pair
operators should have the same mixing amplitudes, i.e., \beq \Gamma^+_n = \sum_{i=1}^{L} y_i P^+_{i1},
~~~\Gamma^+_p = \sum_{i=1}^{L} y_i P^+_{i-1}. \eeq Due to the same reason, the norms for the neutron and proton
wave functions and the matrix elements for the neutron-neutron and proton-proton interaction should satisfy
similar recurrence relations. Therefore below we shall give only the recurrence relations for one kind of
particles, i.e., neutrons. Thus, the norm of the neutron state $|M>$ and the average of neutron number are given
by \beq ~~<M|M> =  \frac{1}{M} \sum_i y_i^2 <M-1|M-1> - \frac{1}{M} \sum_i y^3_i <M-1| P^+_{i1}|M-2> \eeq \beq
<M|N_i|M>= 2 y_i <M| P^+_{i1}|M-1> \eeq where \beq <M|P_{i1}^+|M-1> = y_i <M-1|M-1> - y^2_i <M-1| P^+_{i1}|M-2>.
\eeq

The matrix elements of the neutron-neutron pairing interaction are given by the equations
\begin{eqnarray}
<M|P^+_{i1}P_{j1}|M> &=& y_i^2 y_j^2 <M-2|P^+_{j1}P_{i1}|M-2> \nonumber \\
 &  & + y_j <M|P^+_{i1}|M-1> - y_j^2 y_i <M-1|P^+_{j1}|M-2> \nonumber \\
 &  & + \delta_{ij} y^4_i [<M-2|M-2> - <M-2|N_i|M-2>]
\end{eqnarray}
The iterations are started with the matrix elements $<1|P^+_{i1}P_{j1}|1> = y_i y_j$.
Eqs.(10-13) are very simple and can be used as an alternative to the projecting operator
method commonly applied for systems with like-particle pairing \cite{pbcs}.

The matrix elements of the $T=1$ proton-neutron interaction involve the total wave function $|M M>$. They are given
by the recurrence relation
\begin{eqnarray*}
<M M|P^+_{i0}P_{j0}|M M> =
y_i^2 y_j^2 <M-1 M-1|P^+_{i0}P_{j0}|M-1 M-1>+\\
\delta_{ij} x_i^4 <M-1|M-1>[<M-1|M-1> - <M-1|N_i|M-1>]
\end{eqnarray*}
The starting matrix elements are $<11|P^+_{i0}P_{j0}|11>=\delta_{ij}x_i^4$. As can be seen from the equations
above, the recurrence relations for the PBCS wave functions (2) and (8) are very similar and easy to implement
in numerical calculations.

\section{Results and Discussions}

The results presented in this section correspond to a sequence of $L$ equally spaced 4-fold degenerate levels
(total angular momentum $j=1/2$) with single particle energies $\varepsilon_i=(i-1)/2, i=1,2,...L$ and filled with
$N= L/2$ proton-neutron pairs (quarter filling). We have considered systems with $N=2$ to $N=12$ pairs, which
correspond to typical sizes of open shell $N=Z$ nuclei. The strength of the pairing interaction is varied to cover
all regimes from weak to strong coupling. For these systems we will test the accuracy of the PBCS approximations
comparing correlation energies, odd-even mass differences and occupation probabilities against the exact solution.
We will start this comparison focusing on correlation energies. They are defined as \beq E_{corr}(g) = E_{nor}(g)
- E(g) \eeq where $E_{nor}$ and $E(g)$ are the ground state energies of the system in the normal and in the
correlated phase respectively. Some representative results are shown in Figs. 1-3. All energies are given in units
of the single particle level spacing. In these figures PBC0 corresponds to the variational wave function (2) of
$T_z=0$ $np$ pairs, and PBCS1 corresponds to the variational wave function (8) of $nn$ ($T_z=1$)and $pp$ ($T_z
= -1$) pairs.  The two BCS solutions corresponding to these two types of pairs are called BCS0 and BCS1.
In even-even systems these two BCS solutions are degenerate and are called simply BCS.
Particle number projection breaks this degeneracy.

\begin{figure}
\includegraphics*[height=9cm, width=12cm]{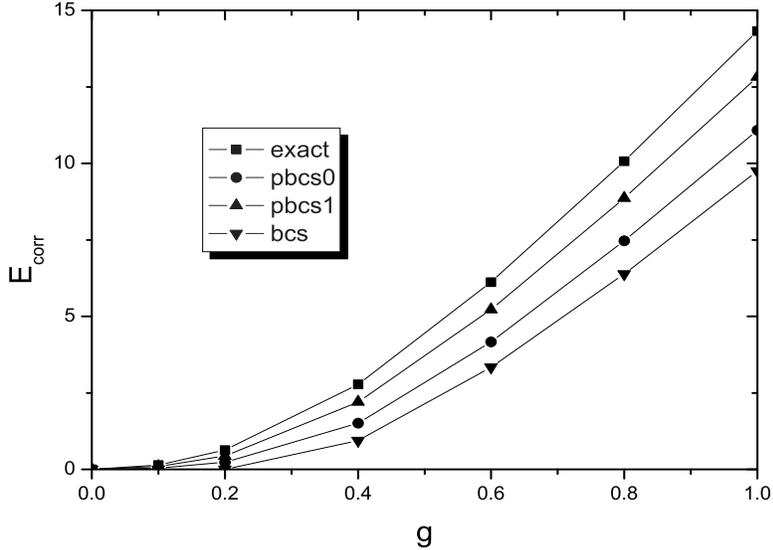}
\caption{ Correlation energy for 4 $pn$ pairs}
\end{figure}
\begin{figure}
\includegraphics*[height=9cm, width=12cm]{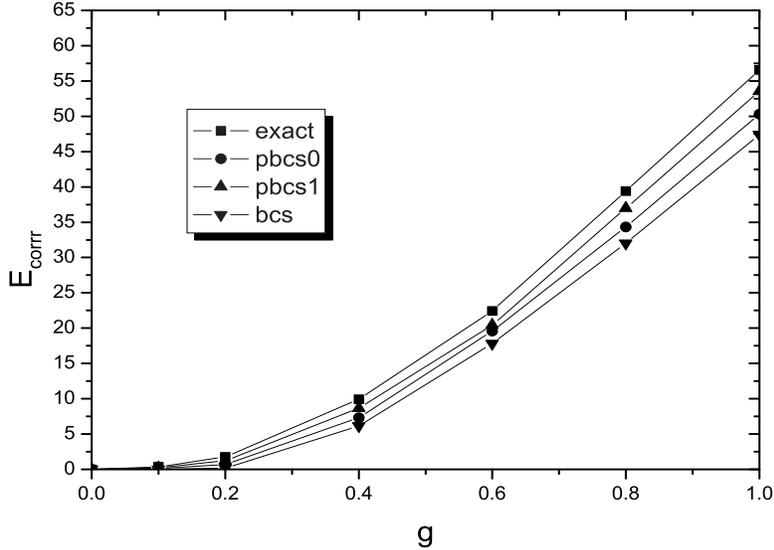}
\caption{ Correlation energy for 8 $pn$ pairs}
\end{figure}

As can be see in Figs. 1 and 2, both PBCS solutions perform better than BCS for even systems, with PBCS1 capturing
more correlations and lowering the ground state energy. On the other hand, as shown in Fig. 3, for a system with an odd
number of pairs the lowest energy solution is PBCS0.
It can be also seen that due to the blocking, in the systems with odd number of pairs the solution PBCS1 becomes
higher in energy even than the BCS solution.

\begin{figure}
\includegraphics*[height=9cm, width=12cm]{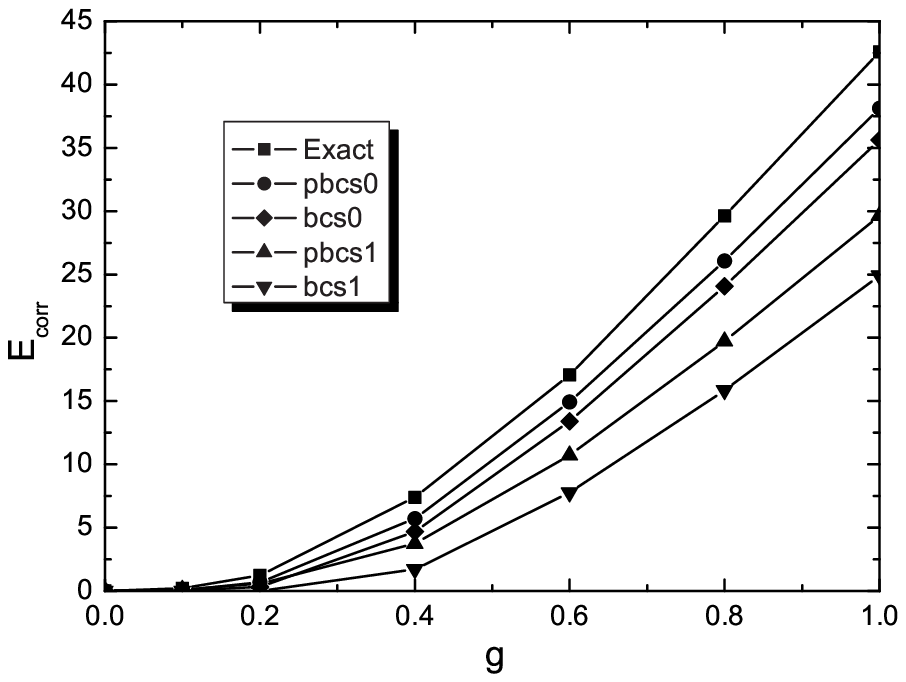}
\caption{ Correlation energy for 7 $pn$ pairs}
\end{figure}

The errors relative to the exact results are shown in Fig. 4. It can be seen that although PBCS gives better
results than BCS, the errors remains significant. The reason is that the PBCS functions (2) and (8) do not
take into account properly the pairing interaction among the pairs with a $T_z$ different from what is
considered in the trial wave function. For example, let's consider the systems with 8 and 7 pairs and 
the interaction strength $g=0.4$. In the system with 8 pairs the wave function PBCS1 gives an energy 
of -25.71 for the 
$T_z = \pm 1$ part of the hamiltonian compared to  -0.84 for the $T_z=0$ part. 
The situation is opposite for the system with 7 pairs: in this case the wave function PBCS0 gives an
interaction energy of -18.04 for the $T_z=0$ component compared to about -1.42 for $T_z = \pm 1$.

\begin{figure}
\includegraphics*[height=9cm, width=12cm]{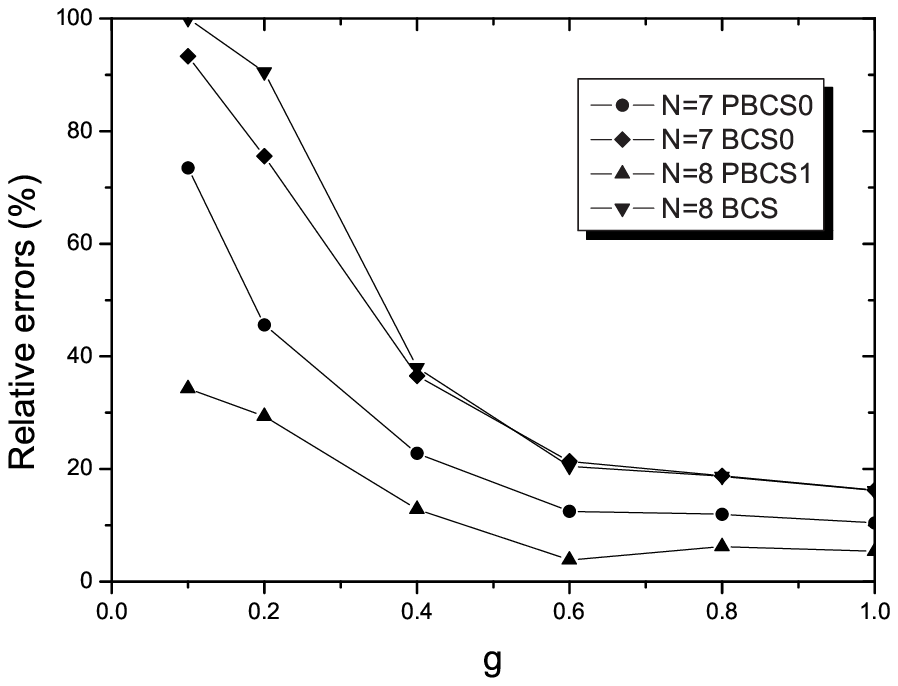}
\caption{ Errors of the correlation energies in systems with 7 and 8 $pn$ pairs}
\end{figure}

Another quantity we have analyzed is the odd-even mass difference along the $N=Z$ line defined as 
\beq
\Delta^{(3)}(M)=\frac{1}{2}[2E(M+1)-E(M)-E(M+2)]. 
\eeq 
Fig. 5 shows the odd-even mass difference for a system
with $M=8$ $pn$ pairs as a function of interaction strength. It can be seen that the PBCS results start to deviate
significantly from the exact values when the interaction becomes stronger. How the odd-even mass difference
depends on the number of pairs is depicted in Fig. 6. As expected, the BCS results do not show the staggering
exhibited by the exact solution. This is because in BCS the solutions (A) and (B) are degenerate in energy. On the
other hand the staggering is present in the PBCS calculations. This is due to the fact that going from the
even-even to odd-odd systems the ground state is changing from PBCS0 to PBC1, which are not degenerate. As seen in
Fig. 6, the shift between the two solutions overestimates the oscillations present in the exact solution. The
reason is that the errors in odd systems are larger than in even systems (see Fig. 4).

\begin{figure}
\includegraphics*[height=9cm, width=12cm]{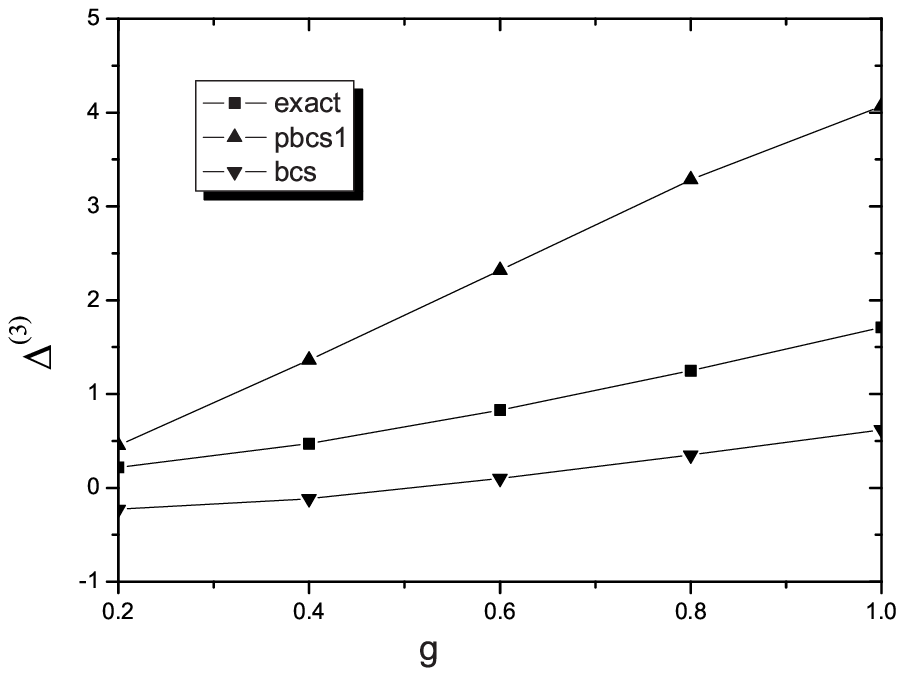}
\caption{ Odd-even mass difference for a system with 8 $pn$ pairs}
\end{figure}
\begin{figure}
\includegraphics*[height=9cm, width=12cm]{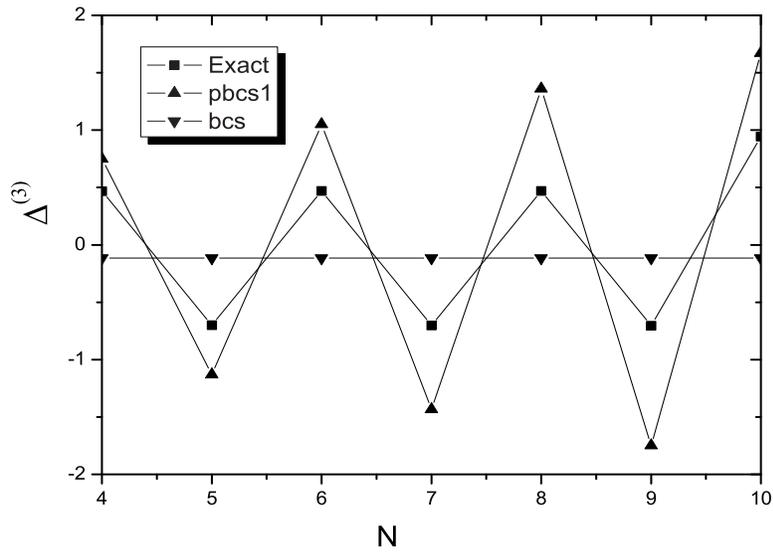}
\caption{ Odd-even mass difference calculated along N=Z line}
\end{figure}

\begin{figure}
\includegraphics*[height=9cm, width=12cm]{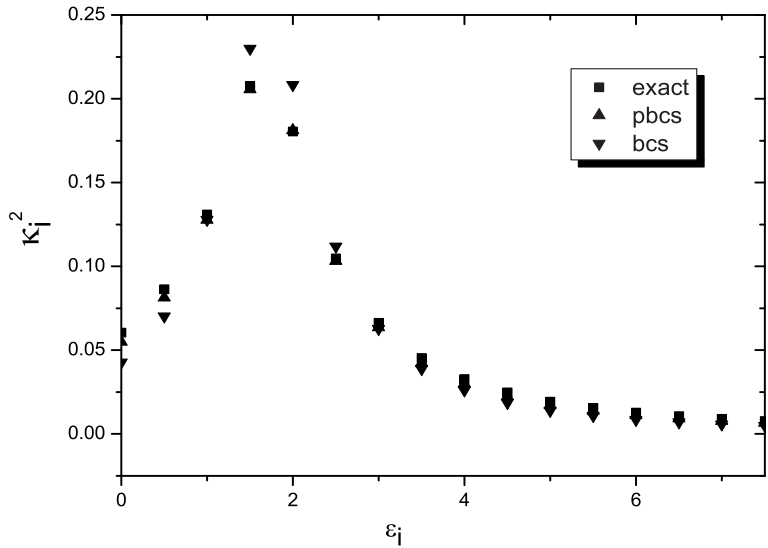}
\caption{ $\kappa_i^2=v^2_i(1-v_i^2) $ as a function of the single-particle energies $\varepsilon_i$
for a system with 8 $pn$ pairs and pairing strength g=0.25}
\end{figure}
\begin{figure}
\includegraphics*[height=9cm, width=12cm]{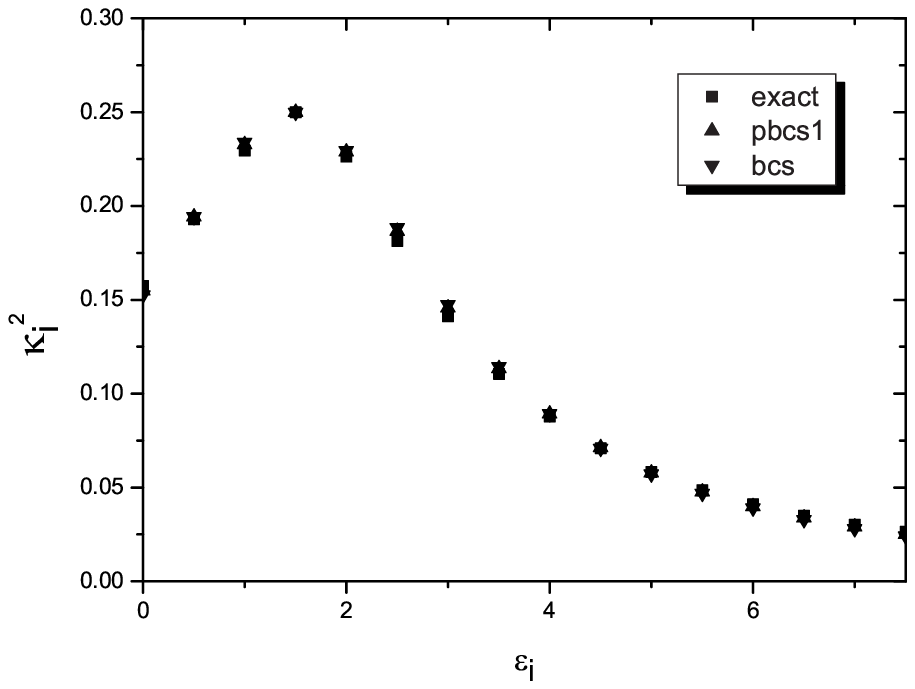}
\caption{The same as in Fig.8 for the pairing strength g=0.4}
\end{figure}

Next we shall discuss shortly the occupation probabilities
corresponding to BCS and PBCS calculations. Figs. 7-8 show the 
quantity $\kappa^2_i=v_i^2 (1-v_i^2)$, where $v^2_i$ is the occupation
probability of the orbit $i$. In BCS $\kappa_i$ is the pairing tensor and
determines the pair transfer form factor. From Figs. 7-8 we can
see that PBCS gives results close to the exact solution for both
values of the coupling strength. 
BCS overestimates the value of $\kappa^2_i$ at the weak coupling (g=0.25) 
in the region around the Fermin energy, where the pairing correlations are
stronger. Conversely, the states further than an energy interval of
the order of the pairing gap are underestimated. These results are
similar to the ones obtained in Ref. \cite{sb} for like-particle
pairing. For stronger interactions (g=0.4) BCS gives results closer to
the exact solution.

Up to now we have considered two distinct PBCS wave functions. The question is if one could get
extra binding by mixing together the wave functions PBCS0 and PBCS1. This is indeed what happens
for a system formed by two $pn$ pairs. The results are shown in the table below. As can be seen,
by mixing the two PBCS states one gets practically the exact result for the correlation energy.
As expected, we get the extra binding when the mixed state has the total isospin equal to  zero.
For systems with more than two pairs a trial wave function with zero isospin cannot be constructed
by mixing only the states PBCS0 and PBCS1.  Consequently for such systems we do not get extra binding
by mixing the two PBCS wave functions.

\begin{table}[hbt]
\caption{ Correlation energies for a system composed of two isovector $pn$ pairs distributed in four
levels with the energies $\epsilon_i=(i-1)/2, i=1,2,3,4$. The binding energies and
the pairing strength $g$ are in units equal to the distance between two
consecutive single-particle levels.}
\begin{center}
\begin{tabular}{c|c|c|c|c}
\hline
9    &     Exact &    Pbcs0+Pbcs1  &    Pbcs1  &  Pbcs0 \\
\hline
0.1 &   0.05587  &     0.0557  &   0.0376 &   0.0189 \\
0.2 & 0.22006 &      0.2192  &   0.1517 &   0.0779 \\
0.4 &   0.81330  &     0.8114  &      0.5924 &   0.3233  \\
0.6 &   1.64761  &     1.6461  &      1.2551 &   0.7364  \\
0.8 &   2.61989  &     2.6190  &      2.0601 &   1.2972  \\
1.0 &   3.66946  &     3.6689  &     2.9487  &   1.9683  \\
\hline
\end{tabular}
\end{center}
\end{table}

\section{Summary and Conclusions}

We have analyzed the accuracy of PBCS approximation for describing isovector pairing correlations in $N=Z$ systems.
The study was done for an exactly solvable hamiltonian with SO(5) symmetry. In the PBCS calculations we
considered two kind of trial wave functions: (1) a condensate of isovector neutron-proton pairs; (2) a  product of
two condensates formed by neutron-neutron and proton-proton pairs. The solution (1) gives the lowest ground state
energy for odd-odd $N=Z$ systems while the solution (2) provides the lowest energy for even-even systems. The PBCS
approximation gives much better correlation energies than BCS, and it is able to describe the staggering of
odd-even mass difference calculated along the $N=Z$ line. However, compared to the pairing between like particles,
for which the PBCS approximation give results very close to the exact  solution of the SU(2) model \cite{sb}, the
accuracy of PBCS approximation for isovector pairing is less satisfactory. The reason is that the PBCS is not able
to treat correctly that part of the isovector force which describes the interaction among the pairs which are not
included in the PBCS condensate. Going beyond PBCS would imply including the isospin projection and/or taking into
account quartet correlations. We are currently working along the later direction.

\vskip 0.5cm \noindent {\bf Acknowledgements} \vskip 0.2cm \noindent We thank R.J.Liotta, P. Schuck and R. Wyss
for valuable discussions This work was supported by Romanian PN II under Grant IDEI nr 270 and by Spanish DGI under
Grant FIS2006-12783-C03-01. B. E. was supported by CE-CAM.

\end{document}